\documentclass[journal]{IEEEtran}
\usepackage{amsmath,amsfonts}
\usepackage{algorithmic}
\usepackage{algorithm}
\usepackage{array}
\usepackage[caption=false,font=footnotesize,labelfont=rm,textfont=rm]{subfig}
\usepackage{textcomp}
\usepackage{stfloats}
\usepackage{url}
\usepackage{verbatim}
\usepackage{graphicx}
\usepackage{makecell}
\usepackage{multirow}
\usepackage{multicol}
\usepackage{threeparttable}
\usepackage{cite}
\usepackage{newtxmath}

\begin{document}
	\bstctlcite{reflist:BSTcontrol}
	
	\title{A Read Margin Enhancement Circuit with \\ Dynamic Bias  Optimization for MRAM}
	\author{Renhe Chen, Albert Lee, Yongqi Hu, and Xufeng Kou,~\IEEEmembership{Senior Member,~IEEE}
		
		\thanks{
			This work is sponsored partly by the National Key R\&D Program of China(Grant No.2021YFA0715503), National Natural Science Foundation of China(GrantNo.92164104), and Xufeng Kou acknowledges the support from the Shanghai Rising-Star program(Grant No.21QA1406000).
			
			Renhe Chen, Zirui Wang, and Xufeng Kou are with the School of Information Science and Technology, ShanghaiTech University, Shanghai 201210, China (e-mail: kouxf@shanghaitech.edu.cn).
			
			Albert Lee and Di Wu are with the Inston Tech, Suzhou 215121, China (e-mail: albertlee@instontech.com, diwu@instontech.com).
			}
		}
	
	
	\maketitle
	
	\begin{abstract}
		This brief introduces a read bias circuit to improve readout yield of magnetic random access memories (MRAMs). A dynamic bias optimization (DBO) circuit is proposed to enable the real-time tracking of the optimal read voltage across process-voltage-temperature (PVT) variations within an MRAM array. It optimizes read performance by adjusting the read bias voltage dynamically for maximum sensing margin. Simulation results on a 28-nm 1Mb MRAM macro show that the tracking accuracy of the proposed DBO circuit remains above 90\% even when the optimal sensing voltage varies up to 50\%. Such dynamic tracking strategy further results in up to two orders of magnitude reduction in the bit error rate with respect to different variations, highlighting its effectiveness in enhancing MRAM performance and reliability.
	\end{abstract}
	

	\section{Introduction}

	\IEEEPARstart{M}{agnetic} random access memory (MRAM) has emerged as a compelling candidate for the next generation non-volatile memory, which offers fast operating speed ($<$10ns), low energy consumption ($<$1pJ/bit), and long endurance ($>$$10^{9}$)\cite{MRAMrev,cali18}. So far, different programming mechanisms, such as spin-transfer torque (STT), spin-orbit torque (SOT), and voltage-controlled magnetic anisotropy (VCMA) effects have been proposed to enhance the energy efficiency of data storage\cite{STT,SOT,VCMAKW}. On the other hand, the read operation of MRAM relies on the tunneling magnetoresistance effect, where the resistance of a magnetic tunnel junction (MTJ) depends on the relative magnetic orientations between the fixed layer and the free layer, as shown in Fig. 1(a). In this context, the data stored in an MRAM cell can be read by comparing the resistance of bit-cell against a reference via a sense amplifier (SA) (Fig. 1(b)).
	
	A major challenge for MRAM read operation is the small tunneling magnetoresistance ratio (TMR) which characterizes the resistance difference between the parallel (P) and anti-parallel (AP) states. This, along with process-voltage-temperature (PVT) variations, results in narrow sensing margins which are inevitably vulnerable to thermal noise and circuit mismatch-induced offset, therefore causing read errors. To address such an issue, calibration modules or offset compensation circuits have been introduced into MRAM sensing circuitry, yet such approaches require additional complexity, area overhead, speed penalty and power consumption\cite{OFFSET_CALI,offset16}. Meanwhile, as for read-disturbance-free MRAMs like SOT and VCMA, sensing margin can be enlarged by raising the readout voltage. However, the TMR ratio normally displays a negative correlation with the bias voltage \cite{TMRV}, and it also changes under temperature and process variations \cite{MTJ_V,TEMP_TMR,TMR200}. Therefore, a higher read bias voltage does not necessarily guarantee a higher sensing margin, and a system capable of dynamically tracking the optimal read bias voltage to maximize the sensing margin hence becomes essential. 

	\begin{figure}[!t]
		\centering
		\includegraphics[width=0.98\linewidth]{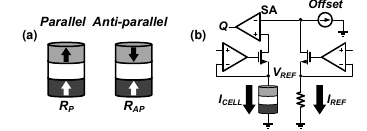}
		\caption{(a) Parallel and anti-parallel states of MTJ. (b) Schematic of current based sensing in MRAM.}
	\end{figure}
	
	In this brief, we utilize a dynamic bias optimizer (DBO) to enhance MRAM readout margin. The remainder of this brief is organized as follows: Section II quantifies the theoretical optimal read bias voltage. Section III provides a detailed explanation of the schematic and operating principle of the DBO circuit. Section IV presents the simulated transient response, tracking accuracy and bit error rate improvement of the DBO-integrated 1Mb MRAM macro. Finally, Section V concludes this brief.
	
	\section{Optimal Bias Voltage for Maximum Margin}
	In general, TMR ratio of an MRAM device at a given temperature can be modeled as\cite{MODEL}
	\begin{equation}
		TMR(V_{REF}) = \frac{TMR(0)}{1+\dfrac{V_{REF}^2}{V_h^2}}
		\label{eqn:tmr}
	\end{equation}
	where $TMR(0)$ represents the TMR ratio under zero bias, $V_h$ is the characteristic voltage at which the TMR ratio drops to half of $TMR(0)$, and $V_{REF}$ is the read bias voltage.
	
	As illustrated in Fig. 1(b), during the MRAM read operation, both the reference and the selected bit-cell are biased at $V_{REF}$, which generate $I_{CELL}$ and $I_{REF}$ for the sense amplifier, respectively. The sensing margin ($I_M$) is defined as the minimum current difference between the cell current and the reference current, that is $I_M=\min\left(I_P-I_{REF},I_{REF}-I_{AP}\right)$. $I_{REF}$ is usually generated by the reference circuit depicted in Fig. 2(a), where the summation of $I_{P}$ and $I_{AP}$ is extracted using two parallel connected MTJ cells in the P and AP states, respectively. Afterwards, the current is scaled by a 2:1 current mirror to generate $\left(I_P+I_{AP}\right)/2$ as the reference current. In this regard, the sensing margin becomes identical for both read 1 and read 0 operations, as expressed by
	\begin{equation}
		\begin{split}
			I_{M}  &= \frac{I_P+I_{AP}}{2}-I_{AP} =I_{P} -\frac{I_P+I_{AP}}{2}\\
			&= \frac{TMR(0)}{2R_P}\frac{1}{\dfrac{1+TMR(0)}{V_{REF}}+\dfrac{V_{REF}}{V_h^2}}
		\end{split}
	\end{equation}	
	\begin{figure}[!t]
		\centering
		\includegraphics[width=0.98\linewidth]{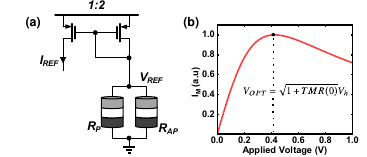}
		\caption{(a) The schematic of an MRAM current reference. (b) Read bias voltage-dependent sensing current margin.}
	\end{figure}
	Accordingly, the read bias voltage dependent $I_M$ curve is shown in Fig. 2(b). Instead of an monotonically increasing correlation, the $I_M-V_{REF}$ curve reaches the maximum value at a certain voltage $V_{OPT}$, and above such an optimal point, any higher bias voltage would lead to a reduced sensing margin instead. Mathematically, $V_{OPT}$ can be derived by taking $\frac{\partial I_M}{\partial V_{REF}}=0$ in (2) as
	\begin{equation}
		V_{OPT}=\sqrt{1+TMR(0)}V_h
		\label{eq3}
	\end{equation}
	
	Guided by \eqref{eq3}, we can observe that the optimal bias point in reference to the largest sensing margin is not a constant. Instead, $V_{OPT}$ is positively correlated with both $TMR(0)$ and $V_{h}$. As summarized in Fig. 3(a), the increase of $TMR(0)$ from 60\% to 140\% can result in a 30\% increase in $V_{OPT}$. Concurrently, by appropriately adjusting the characteristic voltage $V_h$ from 0.15 V to 0.35 V, the optimal bias point is shifted by 100\% (Fig. 3(b)). It is noted that the fabrication process of the MTJ stack invariably causes  discrepancies of $V_h$ and $TMR(0)$ among different memory blocks across the die, and such variations would deteriorate as the MTJ size is scaled down. Therefore, it is important to precisely track the local optimal bias point for individual memory blocks. Moreover, both ambient thermal conditions and circuit heat dissipation would affect the operating temperature of the MRAM array, which also leads to the variations of $TMR(0)$ and $V_h$. For instance, Fig. 4(a) shows that both the $TMR(0)$ and $V_h$ values decrease 30\% when the MTJ bit-cell is heated from room temperature to 400 K, which in turn reduces $V_{OPT}$ by 50\% (Fig. 4(b)). Under such circumstances, real time $V_{OPT}$ tracking is essential for high density, high speed MRAM applications.
	
	\begin{figure}[!t]
		\centering
		\includegraphics[width=0.98\linewidth]{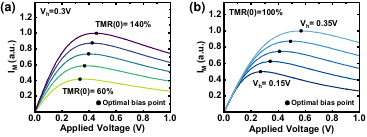}
		\caption{Sensing margin and optimal bias voltage shift at (a) different $TMR(0)$ amplitudes under a given $V_h$ (b) varied $V_h$ for a fixed $TMR(0)$ value.}
		\label{fig:mar}
	\end{figure}
	
	\begin{figure}[!t]
		\centering
				\includegraphics[width=0.98\linewidth]{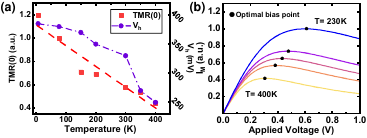}
		\caption{(a) Temperature-dependent $TMR(0)$ and $V_h$ of a typical MTJ bit-cell. Data extracted from\cite{widerange}. (b) Read bias voltage dependent sensing margin curves at different operation temperatures.}
	\end{figure}
	
	\section{Dynamic Bias Optimizer Circuit Design}
	
	\begin{figure*}[!t]
		\centering
				\includegraphics[width=0.95\linewidth]{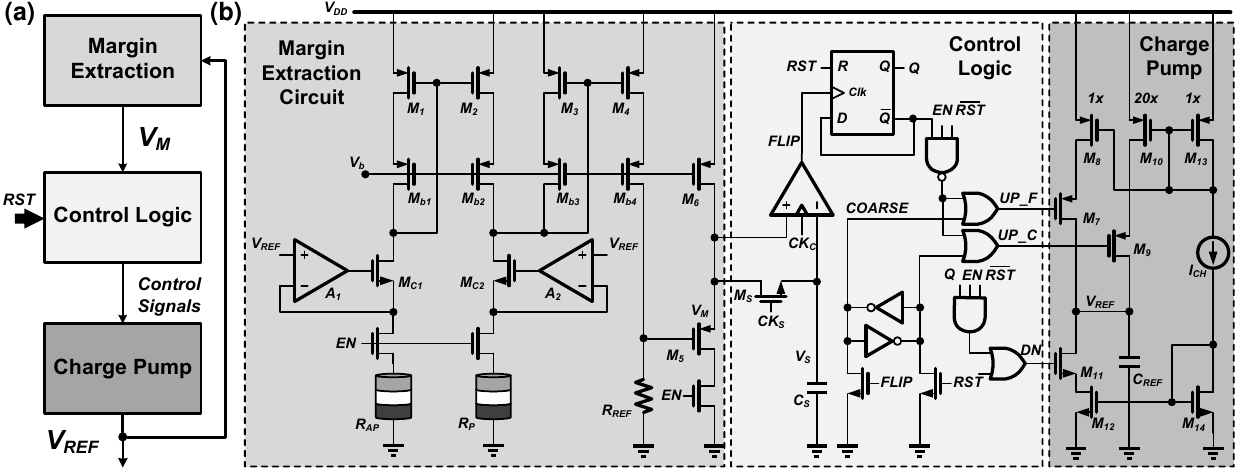}
		\caption{(a) Block diagram of the proposed DBO with feedback mechanism which dynamically tracks sensing margin and generate $V_{REF}$ for MRAM read operations. (b) Detailed schematic of the DBO circuit, which includes margin extraction circuit, control logic block and the charge pump module.}
		\label{fig:DBO}
	\end{figure*}
	
	\begin{figure}[!t]
		
		\includegraphics[width=0.95\linewidth]{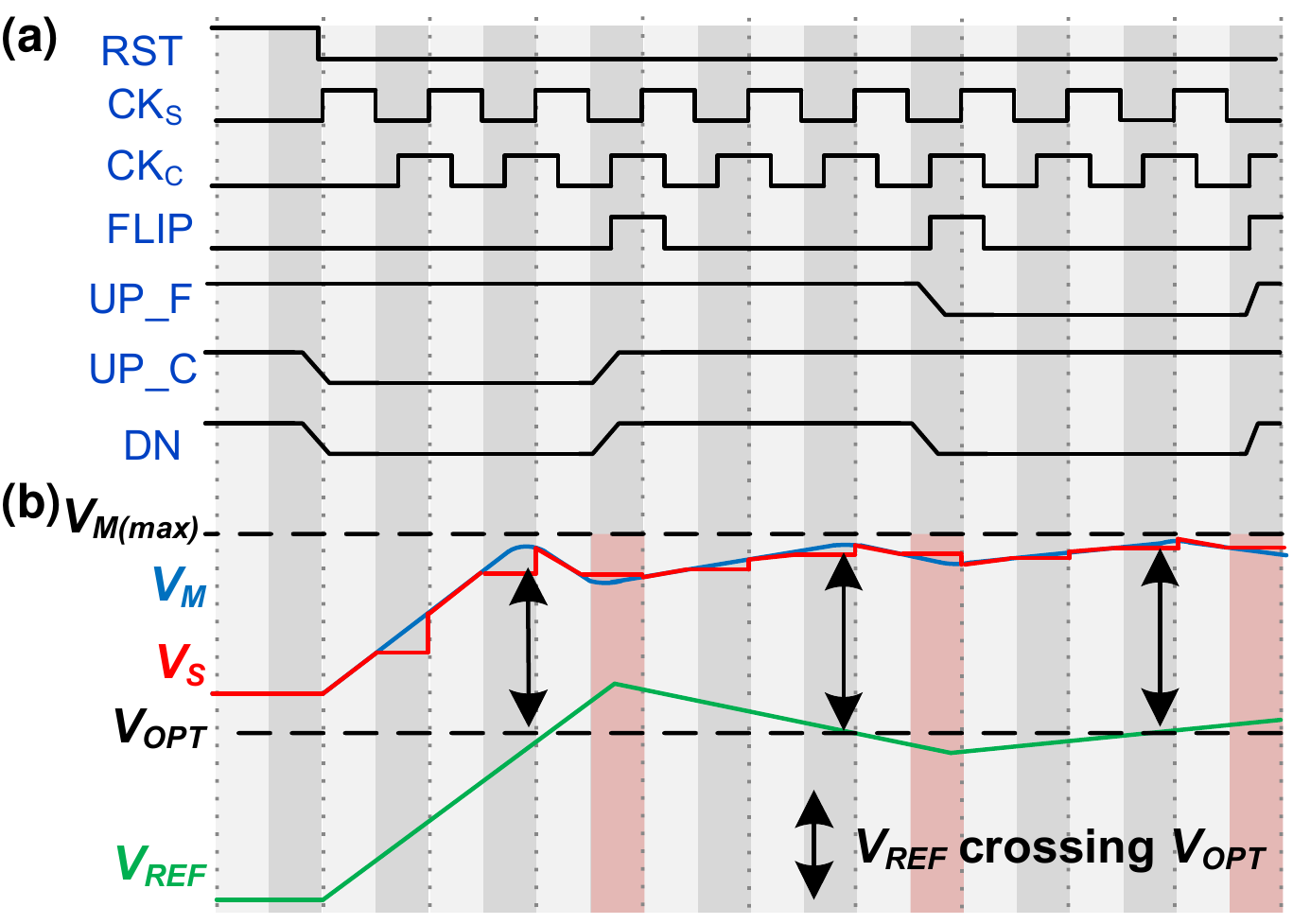}
		\caption{Operation waveforms of the DBO circuit, from initialization to the steady state, with $V_{REF}$ crossing $V_{OPT}$ (marked with arrow) and FLIP events (highlighted with red background).}
		\label{fig:timing}
	\end{figure}
	
	The schematic of the proposed DBO circuit for MRAM read operation is illustrated in \figurename\ \ref{fig:DBO}. In general, the DBO generates reference voltage $V_{REF}$ to approximate $V_{OPT}$ with a feedback system, which consists of three main blocks: 1. a margin extraction circuit that extracts sensing margin $I_M$ and converts it to voltage margin $V_M$, 2. a control logic block that periodically samples $V_M$ to determine the tuning direction of $V_{REF}$, and 3. a charge pump module that feeds back the updated $V_{REF}$ value to the margin extraction circuit as reference voltage in the next cycle.
	
	Specifically, the margin extraction circuit is composed of two active clamp circuits, current mirrors, a trans-impedance amplifier (TIA), and a source follower. To accurately replicate the cell current, MC1/A1 and MC2/A2 regulate the voltage on the reference MTJs to $V_{REF}$, ensuring the current on each path is $I_P=V_{REF}/R_P$ and $I_{AP}=V_{REF}/R_{AP}$, respectively. Meanwhile, the current mirror M1/M2 subtracts $I_{AP}$ from $I_P$ to obtain the sensing current margin $I_M$. Subsequently, $I_M$ is converted to $V_M$ through TIA M3/M4/$\rm R_{REF}$, following $V_M=R_{REF}\times W4/W3\times I_M$. Finally, a source follower isolates the TIA from the control circuits, which not only protects the margin extraction circuit from kickback, but also provides higher driving strength.
	
	Next, in view of the control logic block, it comprises a sample/hold circuit, a comparator, a toggle flip-flop, a latch, and combinational logic. The output voltage $V_M$ from the margin extraction circuit is periodically sampled and stored as $V_S$. The comparator compares $V_S$ and $V_M$ to detect the change of $V_M$, and the FLIP signal determines whether the tuning direction of $V_{REF}$ needs to be adjusted. Additionally, a COARSE flag, set by the FLIP signal, configures the incremental step of $V_{REF}$. In this regard, based on the FLIP and COARSE signals, the control logic generates UP\_C, UP\_F, and DN signals for the charge pump module.

	Finally, a charge pump module updates $V_{REF}$ according to the output signals from the control logic block. In this stage, a bias circuit M13/M14 determines the minimum update step in $V_{REF}$. The current source M7/M8, enabled by UP\_F, provides a fine-tuned increase in $V_{REF}$, while M9/M10, enabled by UP\_C, provides a coarse increase of $V_{REF}$ with 20$\times$ step because of the transistor width ratio between M10 and M8. Conversely, M11/M12, enabled by DN, pulls down the $V_{REF}$ with 1$\times$ step. Therefore, the combination of the coarse-fine paths significantly reduces the settling time during power up and exit from sleep.
	
	Accompanying with the DBO circuit implementation, the corresponding operation waveforms and real-time sensing margin tracking mechanism are visualized in \figurename\ \ref{fig:timing}. Initially, $V_{REF}$ and all latches/flip-flops are reset to 0. Once the DBO is activated, M7/M11 are turned off, while M9 is switched on to enable a fast slewing in $V_{REF}$. In the meantime, the sample/hold circuit samples $V_M$ to $V_S$ at every falling edge of $\rm CK_{S}$. At the end of each hold phase, the comparator that is triggered by $\rm CK_{C}$ would determine the output FLIP state. In particular, as long as $V_{REF}$ approaches $V_{OPT}$, the sensing margin increases (i.e., $V_M>V_S$), and the FLIP signal remains 0 (i.e., indicative of the same charging direction). On the other hand, when the read bias $V_{REF}$ crosses $V_{OPT}$, the sensing margin decreases as $V_{REF}$ deviates away from the optimal bias point, as illustrated in Fig. 2(b). As a result, once the current $V_M$ becomes lower than $V_S$, the FLIP signal is triggered to toggle the charging direction of the charge pump module for the next sampling cycle, and the circuit converges to the optimal $V_{REF}$$=V_{OPT}$ with a periodic toggling of the FLIP signal in the steady state, therefore realizing the real-time tracking of $V_{REF}$ in reference to the $V_{OPT}$ baseline, as shown in Fig. 6(b). It is noted that the first FLIP signal after initialization sets the COARSE flag to 0, hence automatically driving the charge pump stage into the fine-tuning phase to provide a more refined tracking of $V_{OPT}$.
		
	\section{1MB MRAM Simulation}
	
\begin{figure}[!t]
		\centering
		\includegraphics[width=0.95\linewidth]{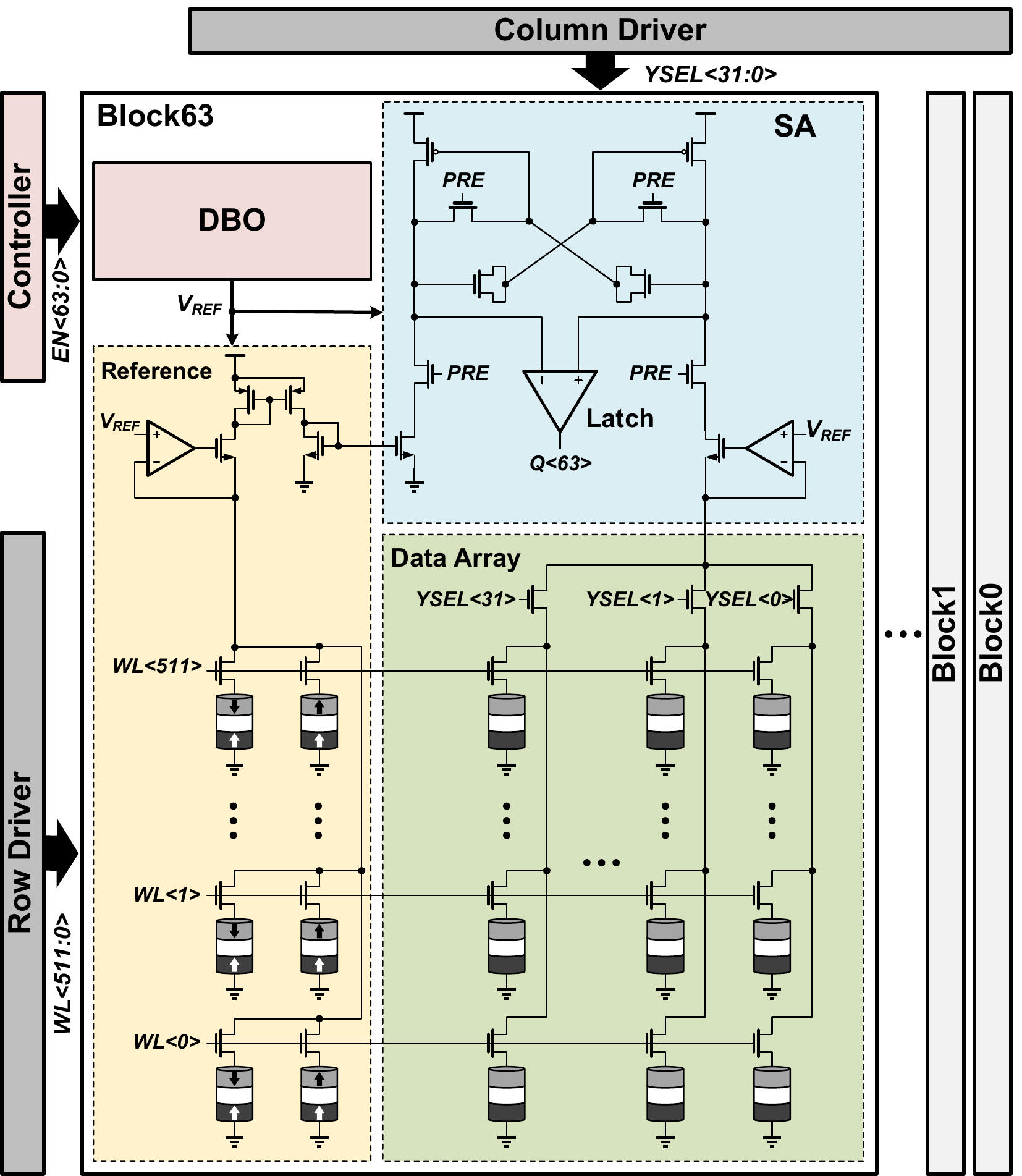}
		\caption{High-level architecture of DBO-enhanced MRAM readout path.}
		\label{fig:HLA}
	\end{figure}

	Following the aforementioned read margin optimization method, an 1Mb MRAM macro with the real-time dynamic bias adjustment function was designed, as depicted in \figurename\ \ref{fig:HLA}. The macro is organized with 64 blocks, each of which consists a 512$\times$34 array with 32 data bit-lines (BL) and 2 reference BLs. Row and column drivers are placed adjacent to the MRAM bit-cell array, which generate WL and YSEL signals for bit-cell selection. To track block level $V_{OPT}$ variation, a DBO module is embedded in each block next to the sense amplifier (SA). It is also seen from \figurename\ \ref{fig:HLA} that the DBOs share cells with the reference columns in the block so as to save layout area. During initialization or exit from sleep, the DBOs help stabilize $V_{REF}$ towards $V_{OPT}$. After stabilization, the DBOs periodically monitor the evolution of $V_{OPT}$ due to temperature induced $TMR(0)$ and $V_h$ drifts, and make dynamic adjustment to $V_{REF}$ accordingly. Besides, each DBO unit is activated individually with 1/64 duty cycle in steady state to reduce power overhead.	
	The performance of the DBO-embedded 1Mb MRAM was evaluated using a 28-nm CMOS process design kit (PDK), and a physical MTJ device model in Verilog-A was used to capture the key parameters from reported measurement results (Table~\ref{tab:table1}) \cite{veriloga}. In addition, we also included the temperature-dependent $TMR(0)$ and $V_h$ data adopted from \cite{widerange} to warrant the reliability of high-temperature simulations.

\begin{figure}[!t]
		\centering
		\includegraphics[width=0.98\linewidth]{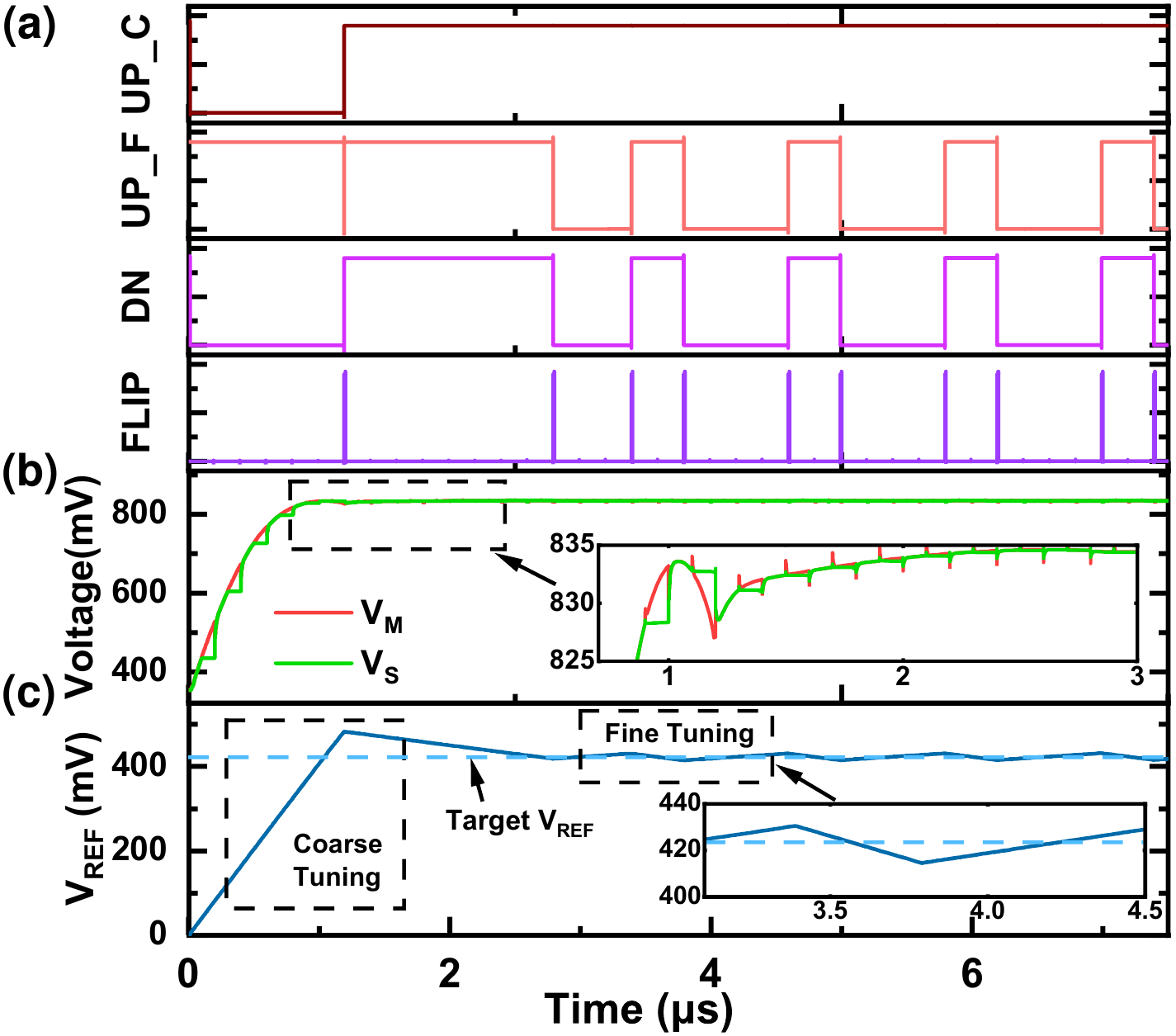}
		\caption{Simulation waveforms of proposed DBO, showing (a) critical control signals (b) $V_M$, $V_S$ and (c) output $V_{REF}$.}
		\label{fig:wave}
	\end{figure}

	 \figurename\ \ref{fig:wave} presents the post-layout simulation results of the designed DBO circuit at room temperature. During the simulation, the sampling rate is set at 5 MHz, and the coarse and fine voltage steps are 80 mV and 4 mV per cycle, respectively. It can be clearly observed from Fig. 8(c) that the $V_{REF}$ curve slews swiftly under the coarse tuning mode during initialization. Afterwards, the control logics promptly responds with FLIP signal pulses whenever $V_{REF}$ crosses $V_{OPT}$. Eventually, the system successfully converges to $(100$\%$\pm2$\%$)V_{OPT}$ within 2µs, or 10 cycles, with a $\pm10$ mV ripple in steady state.
	\begin{table}[!t]
		\caption{\label{tab:table1} Key Parameters Adopted for Simulation}
		\centering
		\begin{tabular}{l l l l}
			\hline
			\hline
			\multicolumn{2}{l}{Parameter}&Description&Value\\
			\hline
			\multicolumn{2}{l}{$\alpha$}&LLGE damping factor&0.02\\
			
			\multicolumn{2}{l}{$M_S$}&Saturation magnetization& 1.2$\times 10^{6}$A/m\\
			
			\multicolumn{2}{l}{$K_i$} & Anisotropy field constant& 1.0$\times 10^{-3}$J/m$^2$ \\
			
			\multicolumn{2}{l}{$X_i$} & VCMA field constant & 80$\times 10^{-15}$ J/(V$\cdot$m)\\
			
			\hline
			
			\multicolumn{2}{l}{$t_{ox}$} & Oxide barrier height & 1.4nm\\
			\multicolumn{2}{l}{$t_{fl}$} & Free layer thickness & 1.1nm\\
			
			\multicolumn{2}{l}{W, L} & MTJ width and length & 60nm\\
			
			\multicolumn{2}{l}{$R_P$} & Parallel resistance & 10k$\Omega$\\
			
			$TMR(0)$ &@RT  & TMR ratio with 0 bias & 100\%\\
			&@125°C &  & 70\%\\
			
			$V_h$ &@RT  & Half TMR voltage & 0.3V\\
			&@125°C &  & 0.22V\\
			\hline
		\end{tabular}	
	\end{table}
	\begin{figure}[!t]
		\centering
		\includegraphics[width=0.98\linewidth]{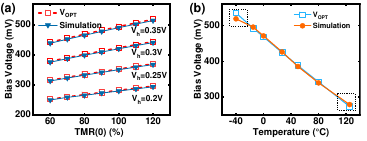}
		\caption{DBO tracking accuracy under different (a) $TMR(0)$ and $V_h$ of MTJ, and (b) thermal conditions.}
	\end{figure}
	
	Apart from the $V_{OPT}$ real-time tracking function validation, the DBO tracking capability under different MTJ electrical parameters and thermal conditions are also evaluated. As illustrated in Fig. 9(a), within the examined range where $TMR(0)$ and $V_h$ vary from 60\% to 120\% and from 0.2 V to 0.35 V respectively, the DBO-adjusted bias voltage $V_{REF}$ manages to follow the evolution of $V_{OPT}$ with the tracking accuracy above 98\%, hence justifying a wide functional range. Likewise, our DBO also exhibits a highly-consistent performance over a wide temperature range from -40 °C to 125 °C, and the $V_{REF}$ value only slightly deviates from the ideal $V_{OPT}$ baseline at extreme high and low temperatures, possibly due to the headroom compression of the current mirrors (M1-M4) and a reduced sensitivity of the comparator in the control logic stage, as shown in Fig. 9(b). Furthermore, by introducing the temperature variation of $TMR(0)$ and $V_h$ during simulation, the transient $V_{REF}$ waveform is able to respond to the rapid temperature drift rate of 98 °C/ms (Fig. 10(a)), and the readout sensing margin improves by 20\% compared to that of the MRAM block without DBO, as highlighted in Fig.10(b). Consequently, the above results evince the advantage of DBO in optimizing the read performance of the MRAM array against process and thermal variations.
	\begin{figure}[!t]
		\centering
		\includegraphics[width=0.95\linewidth]{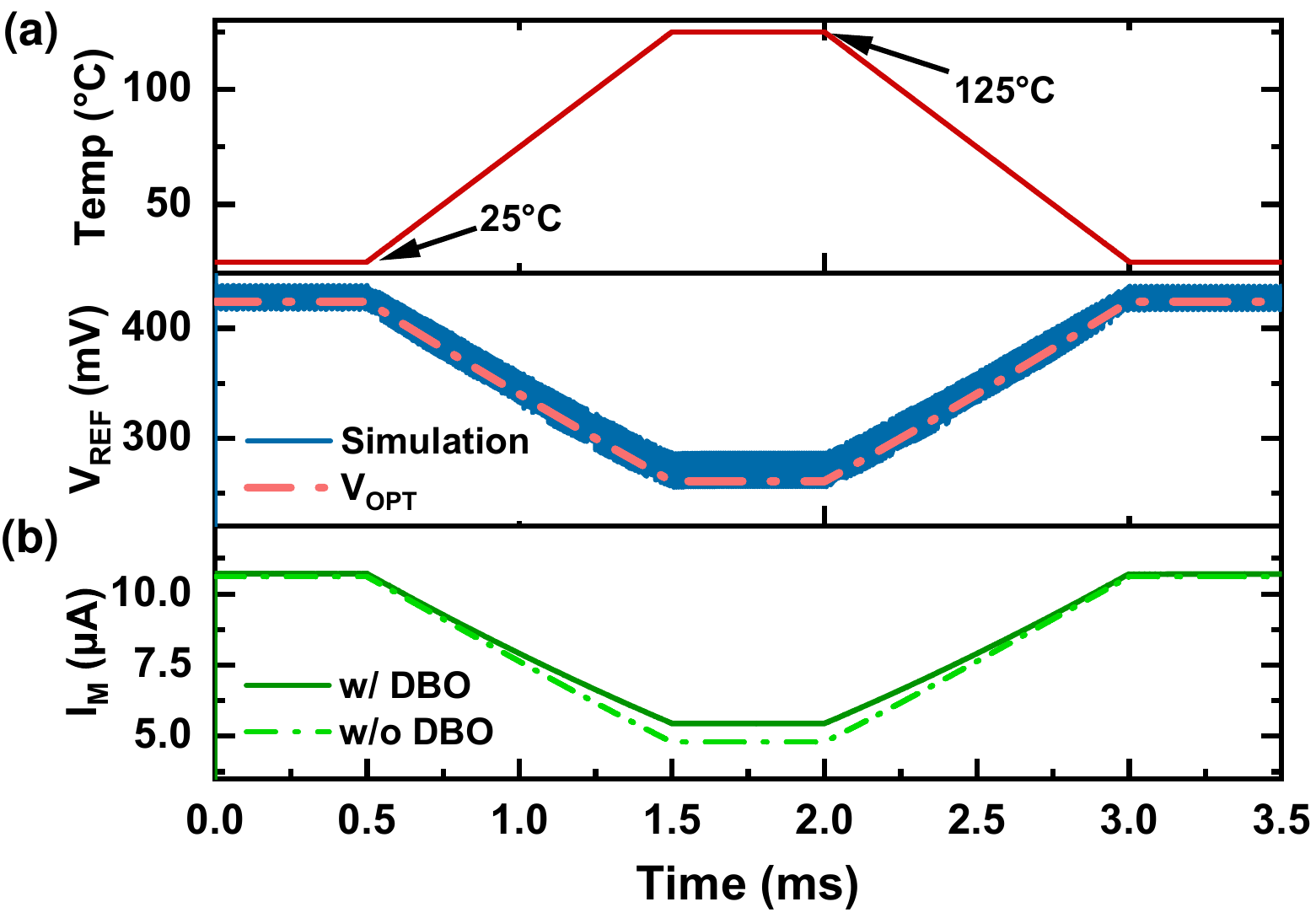}
		\caption{(a) Simulated DBO regulated $V_{REF}$ transient response to rapid temperature change and (b) corresponding sensing margins with (solid line) and without (dashed line) DBO.}
		\label{fig:temptrack}
	\end{figure}
	
	\begin{figure}[!t]
		\centering
				\includegraphics[width=0.98\linewidth]{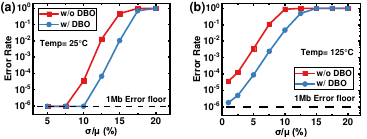}
		\caption{Simulated bit error rate of 1Mb MRAM macro with (blue circles) and without (red squares) DBO versus process variation at (a) $T = 25$ °C, and (b) $T = 125$ °C.}
		\label{fig:yield}
	\end{figure}
	
	Finally, \figurename\ \ref{fig:yield} examines the bit error rate of the 1Mb MRAM macro with (blue circles) and without (red squares) the DBO module by taking into account different degrees of variation-to-mean ratios ($\sigma/\mu$) in $V_h$ and $TMR(0)$. Strikingly, the simulation results at both temperatures consistently confirm that the DBO helps to reduce the bit error rate by 1 to 2 orders of magnitude. From Table II, it can be referred that under the same physical parameters, MRAM array with DBO provides more than 10$\times$ improvement in bit readout accuracy over fixed reference voltages, which is at the expense of a 5\% power overhead and a 6\% extra layout area penalty.
	\begin{table}[!t]
		\caption{\label{tab:table2} 1Mb MRAM Performance Comparison}
		\centering
		\begin{threeparttable}
			\begin{tabular}{c|c|c|c}
				\hline
				\hline
				&w/DBO & \makecell{w/o DBO. \\optimized \\ for 25°C} & \makecell{w/o DBO. \\optimized \\for 125°C} \\
				
				\hline
				{Technology}& \multicolumn{3}{c}{28-nm}\\
				\hline
				{Macro size} & \multicolumn{3}{c}{1Mb}\\
				\hline
				{TMR(0)@25°C [\%]} & \multicolumn{3}{c}{100}\\
				\hline
				{Assumed $\sigma/\mu$ [\%]} & \multicolumn{3}{c}{5}\\
				\hline
				{Operation BER$^*$} &8.71$\times10^{-5}$ & 3.45$\times10^{-3}$&1.27$\times10^{-3}$\\
				\hline
				{Read Power [mW]} & 4.07 & 3.94 & 3.91\\
				\hline
				{Layout Area [mm$^2$]} & 0.242  & 0.229 & 0.229\\
				\hline
			\end{tabular}
			\begin{tablenotes}\footnotesize
				\item[*] Operation bit error rate (BER) is the maximum BER under temperature range from 25 °C to 125 °C
			\end{tablenotes}
		\end{threeparttable}
		
	\end{table}
	\section{Summary}
	In conclusion, the optimal bias condition for the MRAM readout operation is quantitatively investigated and a DBO-modulated $V_{REF}$ is proposed to warrant the largest sensing margin as well as to effectively accommodate the $V_{OPT}$ variation within the MRAM array. The 1Mb MRAM simulation results demonstrate a $>$98\% tracking accuracy regardless of PVT variations. Furthermore, through Monte Carlo simulations, the DBO also brings about 10$\times$ to 100$\times$ improvements in readout yield. Based on the generic feedback principle and margin extraction mechanism, advanced searching algorithm can be applied to the DBO circuit to further improve the $V_{OPT}$ tracking efficiency for high-density MRAM applications.

	\bibliographystyle{IEEEtran}
	\bibliography{reflist}

\end{document}